# Simulate High Traffic and Effective Cost an Academic Kurdistan Network Based on DWDM using OPNET

AYAD GHANY ISMAEEL
Professor
Computer Science PhD in ISE Dept- Technical Engineering College/
Erbil Polytechnic University, Kurdistan
Iraq
RAGHAD Z. YOUSIF
Assistant Professor
Communication Engineering PhD in College of Science
University of Salahaddin, Kurdistan
Iraq

**Abstract:**

Till now the academic institutes in Kurdistan region not connected via network with each other and with research centers worldwide, this paper suggests design an optical IP-Network for academic Kurdistan region based on DWDM (Dense wavelength Division Multiplexer) backbone. The design has been developed using minimum spanning tree between the main realistic locations for campuses in each province were calculated using Prim's algorithm in province level. This proposed network simulated using OPENT IT GURU, and then information was collected using GPS and GIS about the campuses (to obtain realistic locations) of each university in university level, finally unique design model was assumed for each campus in campus level. Simulation the proposed design shows results of delay, traffic sends, traffic receives; utility, performance (packet/second) and throughput are measured in the case of heavy load. Addition to that the hub based network has longer response time than the switch based network.

Ayad Ghany Ismaeel, Raghad Z. Yousif- **Simulate High Traffic and Effective Cost an Academic Kurdistan Network Based on DWDM using OPNET**

## 1. Introduction

In spite the fact that Kurdistan region having more 11,000 staff ,sub staff and postgraduate students but till now the universities of Kurdistan- Iraq doesn't connected by network with each other neither with universities, center research, as well as resources information all over the world. To combat this need, a design is proposed for the Kurdistan academic IP-Network as Intranet. The characteristics of this network must be Hyperlinks (Higher Bandwidth) to implement different application (multimedia, real time, and video conferencing), Due to increased use of computer in classes, labs and faculty offices, the campus network has experienced tremendous growth in the volume of traffic networked that means provide multi-directional information exchanges, and provide same level of services (freely information) to their clients (boundary community) to achieve learning and research objectives.

The suggested design contains campus networks, WAN network, and remote connections, as a first IP-Network in Kurdistan region and Iraq it based on DWDM (Dense Wavelength Division Multiplexing) which is essential for realizing high capacity light wave transport and flexible optical network. The proposed network (Intranet) is based on TCP/IP model (same protocols that apply to the Internet), [1, 2]. When the information send across an Intranet the data is broken into small packets then the packets are sent independently through a series of switches called Routers. The TCP breaks the data into packets and recombines them on the receiving end while the IP handles the routing of the data and makes sure it gets sent to the proper destination [3].

The design proposal depends basically on the switched Ethernet subnets, fast Ethernet, ATM, SONET device with OC-

48 and optical HUB as backbone which is a new technique to ease congestion [2]. The campus network has invested in high-end routers and additional links to the Internet.

## 2. Related Work

Ayad Ghany Ismaeel [2005], proposed design for the Iraqi Universities Intranet Network called IUIN. This Intranet is used for internal corporate communication and included academic Kurdistan region; based on TCP/IP model to employs all applications associated with the Internet such as Web pages, Web browsers, FTP sites, Email, etc [9]. This proposed design; failure to adopt realistic location enough of campuses because not using GPS and GIS to find these locations, not take the academic expansion in Kurdistan region as now, addition there isn't whole simulation for the proposed design but some important parts like server using Little's law, traffic using traffic grooming, etc.

Darwin R. [2009], proposed design for the scientific information network in Kurdistan region, this proposed network simulated using OPNET [10]. Important drawbacks not used the GPS and GIS to determine campuses locations and this proposed IP network somewhat limited in services.

Motivation proposed design of IP-Network can overcome the drawbacks of previous designs.

## 3. Minimum Spanning Tree (MST)

How to select the suitable layout of a network is a very important factor to be discussed, planning to connect the endpoints (nodes) by communications lines. Simply assume that the lines will have sufficient capacity for traffic, and wish to find the least cost network for connection. This type of problem is called a minimum spanning tree problem; there are several methods (or algorithms) for solving this type of problems the popular one is Prim's algorithm which can satisfy high traffic and effective cost [4, 5]: the Prim's Algorithm is described follow:

```
Input: A weighted connected graph G with n vertices and m edges
Output: A Minimum Spanning Tree T for G
Q = new heap-based priority queue
s = a vertex of G {pick up any vertex s of G}
Initialize T to null
for all v ∈ G.vertices()
  if (v = s) then setDistance(v, 0)   {set the key to zero}
  else setDistance(v, ∞)
  setParent(v, Ø)                      {parent edge of each vertex is null}
  Initialize the Q with an item ((u, null), D[u] for each vertex u, where (u, null)
  is the element and D[u] is the key. D[u] is the distance of u
while ¬ Q.isEmpty()
  (u, e) ← Q.removeMin() Add vertex u and edge e to T
  for all e ∈ G.incidentEdges(u)
    z ← G.opposite(u,e)   {for each vertex z adjacent to u such that z is in Q}
    r ← weight(e)         {= w(u, z) }
    if r < getDistance(z)
      setDistance(z, r)    {update the D[z] in Q}
      setParent(z, e)      {update the parent of z in Q}
return the tree T
```

The algorithm is programmed using Java applet. The program GUI is shown in Figure 1, in which the resulted minimum spanning tree is constructed using Prim's algorithm for the proposed network that pass through the main cities in Kurdistan region, in which there are academic institutions will determine the location using GPS and GIS to reach realistic locations, which clear the effect of finding realistic locations of these campuses from not realistic [11].

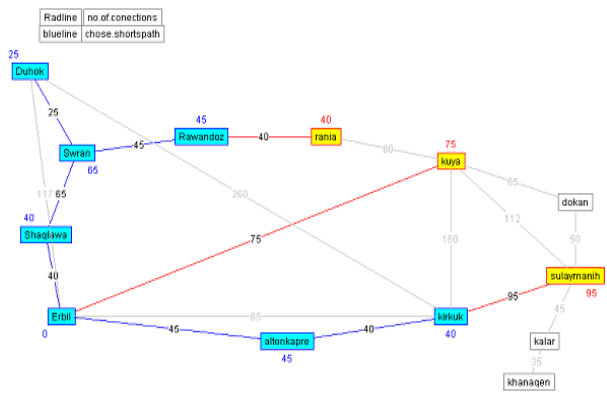

**Figure 1. MST constructed using Prim's algorithm**

Figure 1. shows the start with one of the realistic locations, say Erbil university campus, try to find location that can be connected to it most cheaply, in this example it is Koya university campus. Then find the new location that can be connected most cheaply to either Erbil university campus or Koya university campus, it is Erbil university campus that connected to Kirkuk university campus, after that find the location that can connect most cheaply to it. Then add this link to the tree and continue in this way until all the locations are connected.

## 4. Dense WDM Systems

By the mid-1990s, Dense WDM or DWDM systems were developing with 16 to 40 channels and spacing from 100 to 200 GHz, by the end 1990s DWDM. Systems had evolved of 64 to 160 parallel channels.

In optical network establishment wavelengths are combined onto a single fiber, using DWDM technology several wavelengths, SONET device, or light colors, can simultaneously multiplex signals of 2.5 to 40 Gb/s each over a strand of fiber [4, 6].

By using DWDM as a transport for TDM obtain advantages like SONET multiplexing equipment can be avoided r by interfacing directly to DWDM equipment from ATM and packet switches, where OC-48c (interface is common SONET/SDH compliant client device) see Figure 2.

Ayad Ghany Ismaeel, Raghad Z. Yousif- **Simulate High Traffic and Effective Cost an Academic Kurdistan Network Based on DWDM using OPNET**

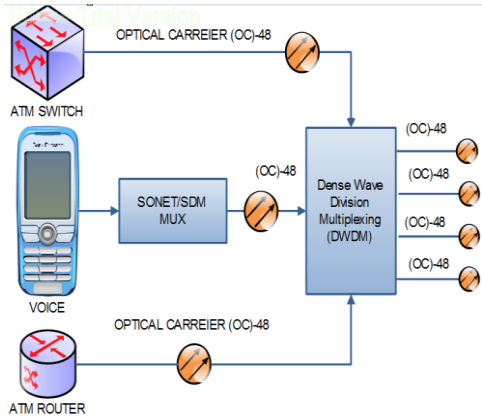

**Figure 2: Direct SONET interfaces from switch to DWDM.**

## 5. Proposed IP-Network Modeling With OPNET

In this section, the stages of our proposal to design the Kurdistan academic IP-Network will illustrate. Our campus is broken down into two parts: Academic and Residential. Academic consisting of academic and administrative buildings. Residential is the housing for students. With some minor exceptions, all the wiring consists of various speeds of fast Ethernet, OC-48 and almost all of the hardware consists of Cisco switches and routers.

Once we discovered the layout of the network and the kind of hardware used, the next step is to see if that hardware existed in OPNET model library [7, 8]. Our network uses the Cisco products, depicted in Table 1. However, that ease-of-use came with a price: Modeler saw all of the ports as being 100/OC-48 enabled.

**Table 1. Reveals Cisco products**

| Cisco device number | Network Level | Note |
|---|---|---|
| ONS 15454 MSTP | DWDM backbone | is optical HUB |
| 7609/7650 | SONET | using with ATM |
| 4500/4700 | router (campuses together) | Get WAN |
| 2900/2924 | campus | Inside one campus |

Academic buildings are not so easy in carbon copies of each other. But at the same time, the idea was the same: Ethernet client machines connected to 2924 with 100Mb copper connections where applicable, 2924's were, in most cases, up linked to 4500's or 4700's using Fast Ethernet. There were other cases where a given 2924 would just be connected directly to the 4500 using fast Ethernet, up linked to 7609's or 7650's using OC-12 is rated to 622.080 Mbps, up linked to 15454 is new technology discovered using OC-48 is rated to 2488.320 Mbps. The "Create LAN" wizard came in handy more times that we could count. The basic target of our design is to make comparing between cost and reliabilities, so, each campus inside contain number of switches of type 2924 connected together by ring topology and one switch is a core of campus connected all to it as shown in Figure 3.

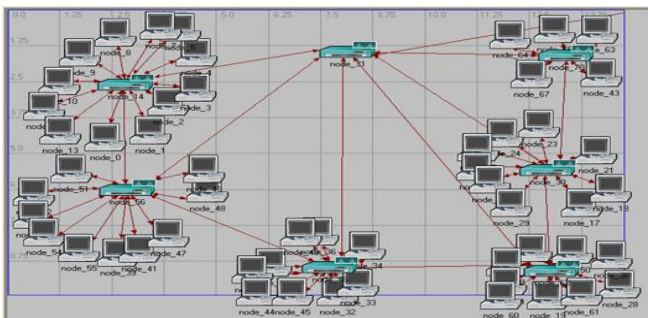

**Figure 3. Proposed Unique Design for Campuses**

Ayad Ghany Ismaeel, Raghad Z. Yousif- **Simulate High Traffic and Effective Cost an Academic Kurdistan Network Based on DWDM using OPNET**

Each campus is connected directly to the SONET optical multiplex, making full real of the Cisco layer number of parts on the (4700) instead of connection each campus to each other like ring network see Figure 4. They use suggest to connect each campus switch to the nearest campus (secondary link) to ensure that the intranet work if fail happed in prim link to the SONET. Finally the SONET device located in small geographical area are connected logical ring to the DWDM backbone of the network by OC-48 as shown in Figure 5.

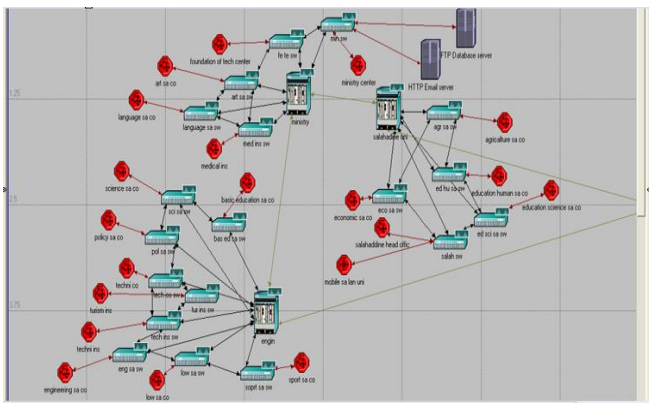

**Figure 4: SONET level with campuses by OC-12.**

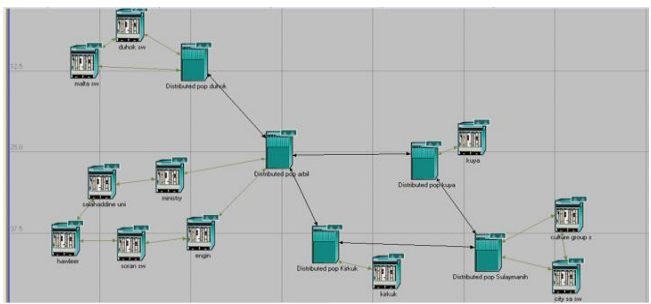

**Figure 5: DWDM Backbone & SONETOC-48**

## 6. Simulation Result

The simulation results for proposal network are including the following item:

Ayad Ghany Ismaeel, Raghad Z. Yousif- **Simulate High Traffic and Effective Cost an Academic Kurdistan Network Based on DWDM using OPNET**

$$Ethernet\ Delay = \frac{\sum x}{t}\ \text{packet/sec,}$$

where x represents the number of packets transmitted at time t

$Ethernet\ Load\ \frac{\sum x_b}{t}$ (bit/sec) ,where $x_b$ represents the number of bits transmitted at time (t). The Ethernet delay is shown in Figure 6 in which blue curve is for switch other red curve is for hub a summary of this Figure is presented in Table 2.

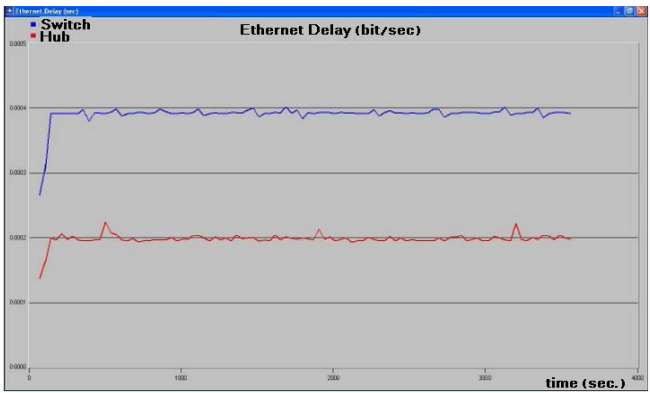

**Figure 6: Ethernet Delay (bit/sec).**

**Table 2. Ethernet delay**

| Ethernet delay | | | | | |
|---|---|---|---|---|---|
| Switch (packet/sec) | | | Hub packet/sec | | |
| Min | Max | Avg. | Min | Max | Avg. |
| 0.000264 | 0.000402 | 0.0003908 | 0.000137 | 0.000225 | 0.000199 |
| Variance | 2.53154053243E-010 | | 7.69492359525E-011 | | |
| Standard deviation | 1.59108156058E-005 | | 8.77207136043E-006 | | |

Figure 7 shows the Ethernet load in which blue curve is for switch other (red) curve is for hub Table 3 shows the summary of these results.

Ayad Ghany Ismaeel, Raghad Z. Yousif- **Simulate High Traffic and Effective Cost an Academic Kurdistan Network Based on DWDM using OPNET**

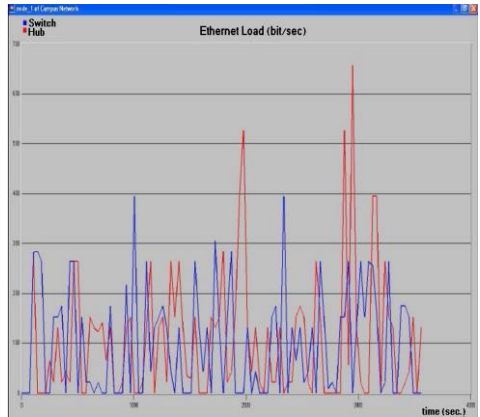

**Figure 7: Ethernet load (bit/sec)**

**Table 3. Ethernet load**

| Ethernet Load (bit/sec) | | | | | |
| --- | --- | --- | --- | --- | --- |
| Switch | | | Hub | | |
| Min | Max | Avg. | Min | Max | Avg. |
| 0.0 | 394.667 | 97.893 | 0.0 | 104.507 | 657.778 |
| Variance | 107.751348329 | | 16.664.671684 | | |
| Standard deviation | 11.610.3530667 | | 129.091718108 | | |

Figure 8 shows Ethernet throughput in which the blue curve is for switch other (red) curve is for hub. Table 4 presents throughput Task network.

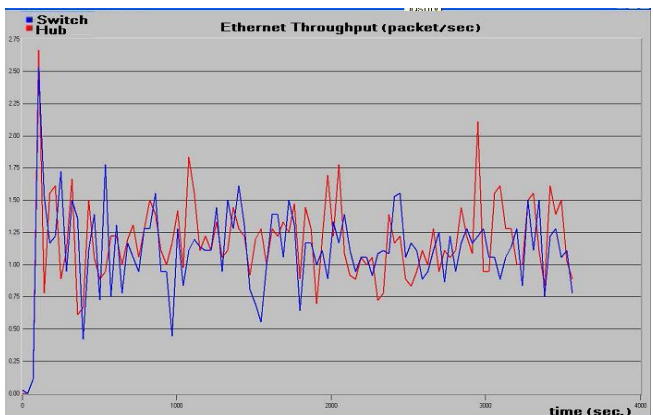

**Figure 8: Ethernet throughput (pack/sec)**

Ayad Ghany Ismaeel, Raghad Z. Yousif- **Simulate High Traffic and Effective Cost an Academic Kurdistan Network Based on DWDM using OPNET**

**Table 4. Throughput**

| | Ethernet Load (packet/sec) | | | | |
|---|---|---|---|---|---|
| Switch | | | Hub | | |
| Min | Max | Avg. | Min | Max | Avg. |
| 0.0 | 2.5278 | 1.1111 | 0.0 | 2.6667 | 1.1675 |
| Variance | 0.119891975309 | | 0.134158256173 | | |
| Standard deviation | 0.346254206196 | | 0.366276202029 | | |

Figure 9 shows the Traffic sent and received (pack/sec).

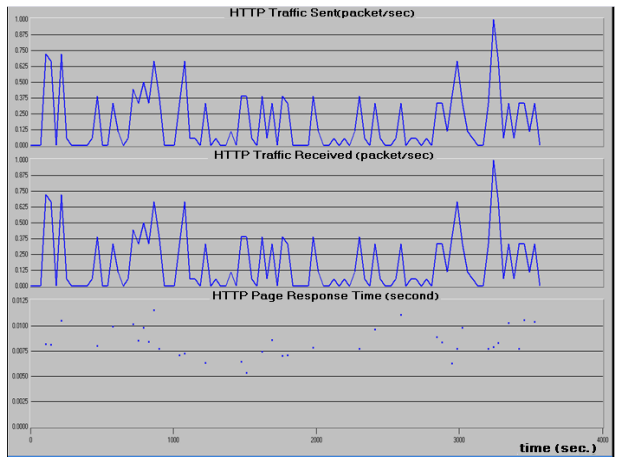

**Figure 9: Traffic sent and received (pack/sec).**

Table 5, reveals comparison between the proposed design and other previous proposed design related.

**Table 5. Reveals comparison the proposed IP-Network with other previous proposed designs**

| The Proposed IP-Network | Design of IUIN [9] | Proposed Design of Darwin [10] |
|---|---|---|
| IP-Network proposed for Kurdistan Academic Institutions only. | All Iraqi Academic Institutions and included Kurdistan Region | IP-Network proposed for Kurdistan Academic Institutions only. |
| Proposed network cover all services and applications include multimedia, voice over IP, videoconference etc. | Unlimited, but not take the academic expansion in Kurdistan region. | Limited in applications and services |

Ayad Ghany Ismaeel, Raghad Z. Yousif- **Simulate High Traffic and Effective Cost an Academic Kurdistan Network Based on DWDM using OPNET**

| | | |
|---|---|---|
| Using GPS and GIS to reach realistic locations of campuses to find the realistic topology of IP-Network using Prim's algorithm [11]. | Unused | Unused |
| Simulated the proposed IP-Network using OPNET shows optimal throughput and performance with unlimited services and applications. | Limited in simulation, and it not used OPNET | Simulation using OPNET |

## 7. Conclusions and Future Work

### 7.1 Conclusions

The OPNET package approves the correctness and scientific eligibility of our network design. From simulation results it's clear that there is no problem with multimedia transmission (voice over IP), and this is expected because the optical communication technology solve this problem. The design proposal takes in consideration the future requirements when the scientific establishment and institution increased in Kurdistan region.

Due to different requirement in our network it can be considered as mixed topology network; because it connected as ring topology in campus level if the have many LANs in one campus while connected as mesh topology when campus contain few number of LANs.

As depicted in simulation results that the switch utilized hub and improve the network performance by reduce delay, load and optimal throughput.

### 7.2 Future Work

The cloud in future closest become the next-generation of grid (IP-Network), and there is commonality between cloud and grid in their vision, architecture and technology, although they differ in aspects such as security, programming model, business model, compute model, data model, applications, and abstractions. As future work employing this proposed IP-Network (grid) to support the cloud computing that provider private data centers which are often centralized in a few locations based on effective technique for allocating servers to support cloud using GPS and GIS [12], i.e. reach to excellent network connections and cheap cost, electrical power, etc.